\documentstyle[aps,preprint]{revtex}
\begin{document}
\begin{center}
\large{\bf Isotope thermometry in nuclear multifragmentation}\\
\vspace*{0.5 true in}
B. K. Agrawal and S. K. Samaddar\\
Saha Institute of Nuclear Physics, 1/AF Bidhannagar, \\Calcutta 700 064, India\\
Tapas Sil and J. N. De\\
Variable Energy Cyclotron Centre, 1/AF Bidhannagar, \\Calcutta 700 064, India
\end{center}
\vspace{0.5 true in}
\begin{abstract}
A systematic study of the effect of fragment$-$fragment interaction, quantum
statistics, $\gamma$-feeding and collective flow is 
made in the extraction of the nuclear
temperature from the double ratio of 
the isotopic yields in the  statistical model
of one-step (Prompt) multifragmentation. Temperature is also
extracted from  the isotope yield ratios  generated in the sequential 
binary-decay model. Comparison of the thermodynamic temperature with the
extracted temperatures for different  isotope ratios 
show  some anomaly in both models which is discussed in the context  
of experimentally measured caloric curves.
\end{abstract}
\newpage
\section{INTRODUCTION}
The response of nuclei to high excitations or temperatures has been a 
subject of intense study both theoretically and experimentally for the 
last  several years. From theoretical 
investigations of hot nuclear matter \cite{kup,jaq,ban}
and also of finite nuclei \cite{bon}, it has been suggested that
the nuclear system may undergo a liquid-gas phase transition at high
temperatures. Recent experimental measurements
of the nuclear caloric curve by the ALADIN Group \cite{poc}
in the $Au+Au$ collisions at 600 $AMeV$ tentatively support
such a conjecture. The key element that enters in
such a surmise is the extraction of the
nuclear temperature that they observed to
be nearly constant in the
excitation energy range of $\sim$ 3 $-$ 10 
$MeV$ per  nucleon beyond which the caloric curve
rises almost linearly with a slope   close
to that of a classical gas. Experimental data from the EOS collaboration
\cite{hau,elli} are also suggestive of critical behavior in nuclei;
here too exact determination of the nuclear temperature has the most
essential role to play.

The temperatures of hot fragmenting systems are generally measured
from the double ratios of isotope multiplicities employing the prescription 
proposed by Albergo {\it et al} \cite{alb}
based on the statistical model of prompt multifragmentation (PM) \cite{ran}.
In arriving at the prescription, several simplifying assumptions 
are made, namely,
(i) the fragments are non-interacting, (ii) the fragments
follow Maxwell-Boltzmann distribution, (iii) they
are formed    in their ground states and 
(iv) all their kinetic energies are in the thermal mode,
i.e. collective flow energy is absent. The effects of
the interaction have later been simulated through an effective
excluded volume interaction \cite{gul}; to our knowledge, the
effect of  fragment-fragment interaction on the isotope ratio
temperature ($T_r$) within the freeze-out configuration has however
not been taken into account. Though it is expected that
at high temperatures and low densities the quantum system would  
behave like a classical Maxwell-Boltzmann system,
the importance of invoking quantum statistics in multifragmentation
has been emphasised by several authors \cite{gul,sub,maj}. The qualitative
effect of quantum statistics is to increase the number of bosons
with respect to fermions at low temperatures and high densities, the
isotope ratio and hence the extracted temperature
$T_r$ might have some sensitivity to the choice of statistics.
The assumption of  the  formation of the fragments 
in their ground states  is an  oversimplification. In general,
the fragments are expected to be formed in various
excited states which are not too short-lived.
These excited fragments  subsequently decay either by
particle or $\gamma$-ray emission. These  side-feeding effects
are shown \cite{gul,kol,bon1,hua} to have an important bearing
on the observed multiplicities and hence
on the 
deduced nuclear temperature. The hot fragmenting nuclear
complex that is formed in nuclear collisions may be compressed
depending on the collision geometry which subsequently
decompresses to the freeze-out configuration  generating
significant amount of collective nuclear flow energy. The 
important role played by collective
flow on the fragmentation pattern has been shown
earlier \cite{pal,des}. Its effect on the nuclear
temperature has only been qualitatively studied by
Shlomo {\it et al} \cite{shl} and found to be
nonnegligible. In a systematic step by step approach,
we explore in this paper the effects of the four approximations
listed earlier on the isotopic temperatures by considering
different isotope double ratios and examine whether
they can be considered as good pointers to the thermodynamic
temperature of the fragmenting system.

The physics of the nuclear multifragmentation is not yet
fully established beyond question. The one-step prompt
break-up (PM) looks a very plausible scenario at high excitations
and the sequential binary decay  (SBD) model \cite{swi,pal1}
may provide a better description of the reaction mechanism at lower excitation.
Both  these processes are thermal in nature. From the inclusive
mass or charge distributions or even the scaling of the multiplicities of
the intermediate mass fragments (IMF),
it is however difficult  \cite{pal2} to discuss the 
relative merits of these two competing 
models. If the SBD model is the more viable model,
say, for the yield of nuclear fragments in nuclear
collisions, then the Albergo prescription of extracting 
nuclear temperature from the double
isotope ratios is called into 
question. One notes that in the SBD
model, there is no unique temperature
but a succession of temperatures till
the nuclear fragments are produced in their  particle stable
configurations. It would still be interesting to know what  values of
temperatures one extracts from double ratios in the SBD model
and whether they can offer some added insight
in the nature of nuclear disassembly.

The paper is organised as follows. In Sec. II, we
briefly outline the PM  and SBD models.
In Sec. III, temperatures calculated from both models
are presented and discussed in the context
of experimental data. The conclusions are drawn in Sec. IV. 

\section{THEORETICAL FRAMEWORK}

The multiplicities of fragments produced in
nuclear collisions are experimentally measured quantities;
the nuclear temperature is a derived entity. In the following,
we outline the  models for fragment production
and relate the nuclear temperature to the fragment yield.

\subsection{Prompt multifragmentation}
A hot nuclear system with $N_0$ neutrons and $Z_0$ protons
may be  formed in nuclear collisions at a temperature
$T_0$ with excitation energy $\epsilon^*$ per particle. It may be
initially compressed  in a volume  smaller than its normal
volume. The compressed matter decompresses
and develops a collective radial flow in addition to thermal excitation.
We still assume that the system evolves in thermodynamic equilibrium and
undergoes multifragmentation after reaching the 
'freeze-out' volume at a temperature $T$ different from $T_0$.
If the time scale involved in  expansion
is larger compared to the equilibration time in the
expanding complex (i.e. the expansion is quasi-static),  
this assumption is not   unjustified .  We further assume
that at the freeze-out volume,  the system reaches
chemical equilibrium.

The expansion of the compressed system may be simulated through a
negative external pressure \cite{pal}. If there was no flow, 
at the freeze-out, the kinetic contribution of the thermal pressure
is generally assumed to be cancelled by
interaction contributions, i.e. the system is 
at equilibrium under zero external pressure.
A positive pressure corresponds to compression of the system; similarly
a negative pressure   would cause decompression. If $P_i$ is  the
internal  partial pressure exerted by the radially  outflowing
fragments of the $i$th species at the surface, the total external
pressure $P$ is then given by $P=-\sum_i 
P_i$. The total thermodynamic potential of the system
at the freeze-out volume is given by \cite{pal,ma}
\begin{equation}
G=E-TS-\sum_{i=1}^{N_s} \mu_i\omega_i + P\Omega,
\end{equation}
where $E$ and $S$ are the internal energy and entropy of the system,
$\Omega= V - V_0$ with $V$ as the freeze-out volume
and $V_0$  the normal nuclear volume of the fragmenting system,
$N_s$ the total number of fragment species, $\mu_i$ the chemical
potential and $\omega_i$  the multiplicity. The occupancy 
of the fragments is obtained
by minimising  the total thermodynamic potential
$G$ and is given by
\begin{equation}
n_i (p_i) = \frac{1}{exp\{(e_i-\mu_i)/T\}\pm 1}
\end{equation}
where ($\pm$ )  sign refers to the fermionic and bosonic
nature of the fragments. The single particle energy $e_i$ is  \cite{pal,de}
\begin{equation}
\label{ei}
e_i=\frac{p_i^2}{2m_i}-B_i+{\cal V}_i -\frac{P_i}{\rho_i}.
\end{equation}
Here $B_i$ refers to
the binding energy, $\rho_i$ is the density of the $i$th fragment
species obtained from the momentum integration of the distribution
function given by eq.(2)  and ${\cal V}_i$ corresponds  to the
single particle potential , evaluated in the 
complementary fragment approximation \cite{gro,pal3}.
It is given by  
\begin{equation}
{\cal V}_i=\frac
{\int  exp[-U_i(R)/T] U_i(R) d^3 R}
{\int exp[-U_i(R)/T] d^3R}
\end{equation}
where $U_i(R)$ is the interaction energy of the fragment with
its complementary at a separation $R$ and 
the integration is over the whole freeze-out volume with
the exclusion of the volume of the complementary 
fragment.
Under chemical equilibrium, the chemical potential of
the $i$th fragment species is 
\begin{equation}
\mu_i=\mu_n N_i +\mu_p Z_i.
\end{equation}
The neutron and proton chemical potentials
$\mu_n$ and $\mu_p$ are obtained from the 
conservation of baryon and charge number, $N_i$ and
$Z_i$ being the number of neutrons and protons in the fragment .
The fragment yield is obtained from the phase-space integration
of the occupancy function and for fermions it is given by
\begin{equation}
\omega_i = \frac{2}{\sqrt{\pi}}\Omega \lambda_i^{-3}
J_{1/2}^{(+)}(\eta_i)\phi_i(T).
\label{f-mul}
\end{equation}
For bosons, the corresponding multiplicity is given by
\begin{equation}
\omega_i=g_0 [e^{-\eta_i}-1]^{-1}
+\frac{2}{\sqrt{\pi}}\Omega \lambda_i^{-3}
J_{1/2}^{(-)}(\eta_i)\phi_i(T).
\label{b-mul}
\end{equation}
In eqs. (\ref{f-mul}) and (\ref{b-mul}), $\eta_i$ is the
fugacity defined as 
\begin{equation}
\eta_i = \frac{\mu_i+ B_i -{\cal V}_i + P_i/\rho_i}{T},
\end{equation}
$\lambda_i= h/\sqrt{2\pi m_i T}$ is the thermal wavelength with
$m_i$ as the mass of the $i$th fragment species and $J_{1/2}^{(\pm)}$
are the Fermi and Bose integrals \cite{path} given by
\begin{equation}
J_{1/2}^{(\pm)}  (\eta) = \int_0^\infty \frac{x^{1/2}dx}
{exp\{(x-\eta)\}\pm 1}.
\end{equation}
The first term on the right hand side of eq. (\ref{b-mul}) gives
the number of condensed bosons,    $g_0$ being their ground state
spin degeneracy. 
  The quanity $\phi_i(T)$  is the 
internal partition function of the fragments
and is defined as
\begin{equation}
\phi_i(T)=\sum_s g_s e^{-\epsilon_s^*(i)/T}
\end{equation}
where $g_s$ is the spin degeneracy of the
excited state $s$ of the cluster with excitation energy $\epsilon_s^*(i)$.
The flow pressure $P_i$  is shown
to be related \cite{pal} to the flow energy $E_i$
of the $i$th species in the form 
\begin{equation}
\frac{P_i}{\rho_i} = C(v_{fi},T) E_i
\end{equation}
where $C$ is dependent on the fragment species, the
temperature and also  on the flow velocity of 
the fragments $v_{fi}$. It is found to be
close to 4 except for very light fragments. 

In the limit $\eta_i << 0$, (which is
true when the density is very low ),  $J_{1/2}^{(+)}(\eta) \rightarrow
\frac{\sqrt\pi}{2}e^\eta$, and then from eq. (\ref{f-mul})  the
yield of the fermion fragments reduces
to 
\begin{equation}
\omega_i = \Omega \lambda_T^{-3} A_i^{3/2} e^{\eta} \phi_i(T)
\label{c-mul}
\end{equation}
where $\lambda_T = h/\sqrt{2\pi m T}$ is the nucleon thermal
wavelength with $m$ as the nucleon  mass and $A_i$ the   mass  number
of the $i$th fragment species. In the same limit,
eq. (\ref{b-mul})  for boson yield reduces  also to eq. 
(\ref{c-mul}). This is also the result
obtained from the classical Maxwell-Boltzmann distribution.

If one chooses  two sets of fragment pairs $(A_1,Z_1)$,
$(A_1',Z_1')$ and $(A_2,Z_2)$, $(A_2',Z_2')$ such
that  $Z_1' = Z_1 +p$, $Z_2' = Z_2+ p$,
$N_1' = N_1 + n$, $N_2'=N_2+n$ where $n$ and $ p$ 
are integers, then from eq. (\ref{c-mul})  it follows that the measured
double ratio $R_2$ of the fragment yields can be
used to determine the temperature of the fragmenting system:
\begin{eqnarray}
R_2 &=&  \frac{\omega(A_1',Z_1')/\omega(A_1,Z_1)}
  {\omega(A_2',Z_2')/\omega(A_2,Z_2)}\nonumber\\
&=& \left  ( \frac{A_1' A_2}{A_1 A_2'}\right )^{3/2}
 \frac{\phi(A_1',Z_1',T)\phi(A_2,Z_2,T)}{\phi(A_1,Z_1,T)\phi(A_2',Z_2',T)}
e^{(\Delta B/T)}e^{-(\Delta {\cal V}/T)} e^{(\Delta F/T)}
\end{eqnarray}
where
\begin{eqnarray}
\Delta B &=& B(A_1',Z_1')-B(A_1,Z_1)+B(A_2,Z_2)-B(A_2',Z_2')\nonumber\\
\Delta {\cal V} &=& {\cal V}(A_1',Z_1')-{\cal V}(A_1,Z_1)+
{\cal V}(A_2,Z_2)-{\cal V}(A_2',Z_2')\nonumber\\
\Delta F &=& C[E(A_1',Z_1')-E(A_1,Z_1)+E(A_2,Z_2)-E(A_2',Z_2')].
\end{eqnarray}
In the limit of low density,  the nuclear part
of single-particle potential  becomes relatively unimportant; further 
choosing  $p=0$ and $n=1$, the Coulomb
contribution to $\Delta{\cal V} $ practically vanishes.

Albergo {\it et al} \cite{alb} further  assumed the fragments to be
formed in their ground states and   they did not consider 
any collective  flow.
Then with   $\Delta F = 0$
and $\Delta {\cal V} = 0$ the
temperature is easily determined from
\begin{equation}
R_2= \left  ( \frac{A_1' A_2}{A_1 A_2'}\right )^{3/2}
 \frac{g_0(A_1',Z_1',T)g_0(A_2,Z_2,T)}{g_0(A_1,Z_1,T)g_0(A_2',Z_2',T)}
e^{(\Delta B/T)}
\end{equation}
since the ground state degeneracy $g_0(A,Z)$  and binding
energies are   a-priori known.

If prompt multifragmentation is the real physical
mechanism for fragment production, eq. (15) then  
provides  an  approximate but simple
way to find out the thermodynamic temperature
of the disassembling system.  Influences
from other effects as already mentioned are however 
embedded in the experimental data  for isotope
yield ratios. One can not obtain
informations on the perturbations caused by these
effects on the double-ratio thermometer simply from
the experimental isotopic yields without the
help of further model calculations. If there were
no other effects except from side-feeding  through
$\gamma$-decay, the experimental data could
be exploited to delineate side-feeding effects by
using eq. (13) with $\Delta{\cal V} = 0$ and $\Delta F=0$ with
the choice of the internal partition function
from eq. (10). Effects from particle decay \cite{kol}  or those
coming from  the inclusion of Coulomb force for yield ratios 
involving  isotopes differing by proton number \cite{kolo} could also 
be  approximately reconstructed from the
experimental fragment multiplicities. Influence
of nuclear interaction, quantum statistics or collective
expansion can not however  be singled  out
without recourse to models. We have therefore
done calculations in the prompt multifragmentation
model with the barest scenario (classical statistics,
no interaction, no side-feeding and no nuclear flow) and then
included  the said effects step by step to generate
fragment multiplicities. The multiplicities
so generated under different approximations are used 
to extract double-ratio temperatures using eq. (15) 
to delineate  the role of various effects on the temperatures.
\subsection{Sequential binary decay}

Fragmentation may also proceed via a sequence
of binary  fission-like events, particulary at 
relatively lower excitation energies. We employ the transition-state
model of Swiatecki \cite{swi} to find the decay probability of a
hot nucleus with mass $A$, charge $Z$ and excitation energy $E^*$ into two
fragments of mass and charge  $(A_1,Z_1) $ and 
$(A- A_1, Z - Z_1)$ respectively. At the saddle point, the binary
fragmentation probability is given by 
\begin{equation}
P(A,Z,E^*; A_1, Z_1)\propto  exp\left [ 2\sqrt{a (E^*-V_B-K)}-2\sqrt{a E^*}\right ]
\end{equation}
where $a$ is the level density parameter taken as
$A/10$ $MeV^{-1}$, $K$ is the relative kinetic energy at
the saddle point and $V_B$ the barrier height dependent on
the saddle point temperature $T_s$ which is different from the
temperature $T_0$ of the parent nucleus given by $T_0=\sqrt{E^*/a}$.
The barrier height is determined in the two sphere approximation
as 
\begin{equation}
V_B(T_s) = V_c + V_N + E_{sep}(T_0, T_s)
\label{vb}
\end{equation}
where $E_{sep}$ is the separation energy. It is
evaluated as 
\begin{equation}
E_{sep}(T_0,T_s)= B(T_0)-B_1(T_s)-B_2(T_s).
\end{equation}
The binding energies are taken to be temperature dependent \cite{pal3}.
The saddle-point temperature which is also the 
temperature of the fragmented daughter nuclei is  given as 
\begin{equation}
T_s = \sqrt{(E^* - V_B - K)/a}.
\label{ts}
\end{equation}
The evaluation of $T_s$ from eq. (\ref{ts})  requires a
knowledge of the relative kinetic energy $K$.
We assume it to follow a thermal distribution 
$P(K) \propto \sqrt{K} e^{-K/T_s}$.
The complicated interrelationship between $V_B$, $K$ and $T_s$ 
renders evaluation of $T_s$ difficult; to simplify the problem,
$K$ in eq. (\ref{ts}) is replaced by its average value $\frac{3}{2}T_s$
and then $T_s$ is evaluated in an iterative procedure with
$T_0$  as the starting value. This  is expected
to be a good approximation since the dispersion in   kinetic
energy is of the $\sim$ $T_s$ and $(E^*-V_B)$ is generally much greater than
$T_s$. The so extracted value of $T_s$ is used only to evaluate the
barrier $V_B$ from eq. (\ref{vb}), the decay probability 
and the thermal distribuion.  In eq. (\ref{vb}), $V_c$ is the
Coulomb interaction taken to be that between two uniformly charged
spheres and $V_N$ is the interfragment nuclear interaction \cite{pal3}.

The relative kinetic energy  $K$ of the two fragmented nuclei lying in the
range $0\le K \le (E^*-V_B)$ is generated  in a Monte-Carlo
method obeying the thermal distribution as mentioned. To ensure  energy conservation,
this kinetic energy is plugged into eq. (\ref{ts}) to evaluate
the temperature of the daughter nuclei for further dynamical evolution. The fragment
kinetic energy  and hence their velocities are obtained from momentum 
conservation.

The trajectories of the fragments are calculated
under the influence of Coulomb interaction in the
overall centre of mass frame. If the fragments have sufficient excitation energy,
they decay in flight. The integration  of the trajectories 
is continued till the asymptotic   region is reached when the interaction
energy is very small ($\sim$ 1 $MeV$) and the excitation energy of the
fragments are below particle emission threshold.

\section{RESULTS AND DISCUSSIONS}
 In this section we present the results of the
calculations for temperatures extracted from  double ratios
of different isotope yields obtained from nuclear
multifragmentation. These calculations are performed under
different approximations mentioned in the introduction  in the
PM  model. For this purpose  we have taken
$^{150}Sm$ as a representative case for the fragmenting system. We also
obtained  the double ratio temperatures assuming that the  fragmentation
proceeds   via sequential binary decay.

\subsection{Prompt multifragmentation}
The initial temperature $T_0$ of the hot system formed
is different from the kinetic temperature $T$ 
(also referred to as the thermodynamic temperature) of the fragments
at the freeze-out. What remains constant is the
total energy $E$ of the system or equivalently its
excitation energy $E^*= E+B_0$ where $B_0$ is the
binding energy of the system. The total energy of the fragmented system
may be written as 
\begin{equation}
E=\frac{3}{2}T (M-1) - \sum_{i=1}^{N_s}
\omega_i B_i - \frac{1}{2} \sum
\omega_i {\cal V}_i + \sum \omega_i<\epsilon^*(i)>,
\end{equation}
where $M=\sum_i\omega_i$  is the total number of the
fragments produced in the grand canonical
model for PM and ${\cal V}_i$ the single-particle potential.
The quantity $<\epsilon^*(i)>$ is the average excitation energy
of the $i$th fragment species given by
\begin{equation}
<\epsilon^*(i)> =\frac{ \int \epsilon \rho_i(\epsilon)e^{-\epsilon/T}
d\epsilon}{\int \rho_i(\epsilon)e^{-\epsilon/T}d\epsilon}
\end{equation}
where the integration extend upto  the particle 
emission threshold  and $\rho_i$  is the level density
obtained  from Bethe ansatz\cite{pal1}.
To compare  the temperature $T$ and $T_0$ taken as $T_0= 
\sqrt{E^*/a}$ , we plot in fig. 1 these 
temperatures  as a function of $\epsilon^* = E^*/A$, the
excitation energy per particle.  The dashed line 
corresponds to the temperature $T_0$ and the solid
and dot-dash lines correspond   to the thermodynamic temperatures
evaluated at the freeze-out volumes  6$V_0$ and 10$V_0$
respectively, $V_0$ being the normal volume of the fragmenting system.
The curve for $T_0$ is parabolic but it is
interesting to note that the caloric curves corresponding to
the different freeze-out volumes mentioned show plateaux 
in the excitation energy . In the
canonical model  of multifragmentation with multiplicity-dependent
freeze-out volume, Bondorf {\it et al} \cite{bon}  reported first
such a plateau  reminiscent  of the onset of a phase
transition in nuclei. With increase   in freeze-out volume,
we find in our calculation that  the temperature decreases and the
plateau gets extended in the excitation energy.
Such a dependence of caloric curve on the freeze-out volume was
also observed  in a self-consistent  Thomas-Fermi 
calculation \cite{de1}.

In figures 2-7, we display the isotope double
ratio temperatures $T_r$ from the prompt break-up
of $^{150}Sm$ with different choices of  isotope combinations
fixing the freeze-out volume at $6V_0$. The combinations are:
$(^{4}He/^{3}He)/(d/p)$, $(^{4}He/^{3}He)/(t/d)$, 
 $(^{7}Li/^{6}Li)/(d/p)$, 
 $(^{7}Li/^{6}Li)/(^{4}He/^{3}He)$, 
 $(^{10}Be/^{9}Be)/(^{4}He/^{3}He)$ and
 $(^{13}C/^{12}C)/(^{7}Li$ $/^{6}Li)$. They would be referred to as
$(He-d)$, $(He-t)$, $ (Li-d)$, $(Li-He)$, $(Be-He)$ 
and $(C-Li)$ respectively. 
In all these figures, the dotted lines correspond to the
temperatures obtained from the multiplicities generated in
the barest Albergo prescription as mentioned earlier.
It is obvious that the thermodynamic temperature and the
double-ratio temperatures are identical in this case.
The dashed lines
($V_{int}$ ) refer to the temperatures calculated from
eq.(15) but with the inclusion of
final state interaction (nuclear+Coulomb) over the 
barest scenario for the fragment generation.  In all the
cases investigated ,  it is found that the inclusion of
fragment-fragment interaction (${\cal V}$) shifts the temperature by
nearly a constant amount at all excitations; the
amount of shift or its sign depends on the particular  isotope combination 
chosen. The shift is found to be negligible for double ratios
($He-d$),$(Li-He)$ 
and $(Be-He)$. The dot-dash lines (QS) in the
figures refer to  calculations done with further inclusion of quantum statistics.
As comparison of the dashed and dot-dash curves shows,  
no appreciable quantum
effects  are evident except in the case of 
the temperature obtained from the
double ratios $(Li-d)$. In this particular case,
it is further seen that the difference between the quantum and classical
(Maxwell-Boltzmann) calculations widens  with  excitation energy or with 
temperature. It is normally expected that at
low density and high temperature  \cite{maj}, quantum
effects would not be discernible, to be more exact, as explained
earlier it depends on  whether the fugacity $\eta << 0$. It is seen that
the densities of the fragment species
or alternatively  their fugacity $\eta$ vary in a complex
way with the temperature. When the temperatue
is low, the density is extremely low and hence the value of $\eta$
is relatively large and negative;  with increase in temperature
along with density  the value of $\eta$ increases initially and then
again decreases for the complex fragments. However for nucleons
$\eta$   increases monotonically  in the
energy regime that we consider. This complex 
variation of $\eta$ is reflected in the temperatures
extracted  from the double ratio of yields obtained with quantum statistics.

In order to take into account effects due to side-feeding, we next
assume that the fragments are produced in particle-stable excited
states so that the ground state population from the 
$\gamma$-decaying states have to be considered. Side-feeding
from particle decay is thus ignored. Kolomiets {\it et al}
\cite{kol} have shown that particle-decay effects are rather
negligible, further there is uncertainty about the cut-off
limit to the particle decay width $\Gamma$ that one should
take which is intimately coupled with the time scale for
prompt multifragmentation. Side-feeding effects are studied
after generating the fragment yield by using eqs.(6),(7) and
(8) with flow pressure $P=0$. In these equations, $\phi$ is
the internal partition function that includes a sum extending
over the ground and $\gamma$-decaying excited states.
For the fragments  considered,  isotopes  upto $^4He$ were taken as  
billiard balls with no internal excitation as it has no low-lying 
$\gamma$-decaying state. 
Similarly for $^9Be$, only the ground state was considered. 
For the rest, the excited states considered are 3.563 $MeV$ 
for $^6Li$, 0.478 $MeV$ for $^7Li$, 3.37, 5.958,
5.960, 6.018 and 6.26 MeV for $^{10}Be$, 4.439 for $^{12}C$
and  3.089, 3.685 and 3.854 $MeV$ for 
$^{13}C$. For other heavier nuclei,  continuum approximation is used for the
single-particle levels  and internal partition function is  taken as
$\phi=\int \rho(\epsilon)e^{-\epsilon/T} d\epsilon$ where the integration 
extends upto particle emission threshold.
Over and above the quantum statistical
effects, when we consider effects due to $\gamma$-feeding, it is
found from figs. 4-7 ( by comparing the dot-dash and the full lines)
that these effects are very sizeable. 
The $(He-d)$ and 
$(He-t)$ thermometers show no side-feeding
effects (figs. 2 and 3) as these fragments are taken to have
no excited states. A dramatic effect is seen for the 
$(Be-He)$ thermometer ( displayed
in fig.6) 
where the sharply upward going full line refers to the temperature 
$T_r$  obtained this way. Bondorf {\it et al} \cite{bon}  found a similar
behaviour for the $Be-He$ thermometer.

In central  or near-central   collisions between medium-heavy
or heavy  nuclei at intermediate  or higher energies, compression
and eventual decompression of the nuclear matter manifests itself
in nuclear collective flow energy which might be a significant part of 
the total excitation. Collective flow influences  the
multifragmentation pattern  to a  significant 
extent \cite{pal,des}.  The double-ratio isotope 
thermometer may then need to be recalibrated a great deal due
to the nuclear flow.  This is manifest from the
figures  2 - 7 where the full line  with crosses
correspond to  calculated temperatures with inclusion of flow 
above the effects induced by fragment-fragment interaction, quantum
statistics  and whereever applicable, $\gamma$-feeding.  
The flow energy is taken to be
$25\%$ of the total excitation energy. Comparison of the full
line with  the line with crosses shows  that at a given 
excitation energy, the temperature is always
lower or for the same temperature, the
excitation energy is always higher. In fig. 8, all the
double-ratio isotope thermometers except for Be-He are displayed
for comparison. Except for the
flow effects, other effects are included here.
The behaviour of the temperature profiles 
with excitation energy  look nearly the same
but their magnitudes differ depending on the choice of 
the thermometers. 
At lower excitations, an uncertainty in the $T_r$ $\sim 2.0$ $MeV$
involving $(Li-d)$ and $(Li-He)$ thermometers
is found which increases  progressively with
excitation energy. The uncertainty involving $(He-t)$,
$(He-d)$ and $(C-Li)$ thermometers however decreases
with excitation energy, all three temperatures converging
at the highest excitation we study. 
In fig. 9, 
the isotope temperature corresponding to $(Li-He)$ 
is shown with inclusion of different magnitudes of flow.
The full and dashed curves refer to cases  when half ($50\%$) and 
one fourth ($25\%$) of the total excitation 
have gone to the flow energy; the dotted
curve corresponds to  no flow. As an illustration,
data from the ALADIN \cite{poc} and EOS \cite{hau} experiments
are displayed in the figure, which use the $(Li-He)$ and
$(He-t)$ thermometers respectively. To have a contact with
the EOS data, we also display the calculated temperature
from the $(He-t)$ thermometer with 50$\%$ flow
energy (dot-dash curve). In an analysis   
of the same data  in Ref \cite{sam}, it  was
pointed out that  the data could be better explained 
invoking  progressive increase    of the percentage  of flow
energy with increasing  total excitation;  comparison of the
present calculations with the experimental data   validates this
observation. 

\subsection{Sequential binary decay}

Hot nuclear systems may release  energy through binary fission-like
decay, the decay chain continues till there is no further energy for  binary division.
At the end of such decay process,  fragments of different species  are produced
in ground states and  in $\gamma$-decaying excited states, 
the multiplicity depending on the initial system and excitation energy.
It has been noted earlier \cite{lop} that the frequency distribution of the
fragments follows almost a power-law distribution
and that it is not too different from the
one obtained from  prompt multifragmentation at the same
excitation energy. Our calculations done at different excitation
energies also show  that the inclusive
mass or charge distrbutions obtained from both PM
and SBD models are roughly the same.
The isotopic distributions  are however   seen to have significant  differences.
In the SBD model, the hot nucleus  prepared initially at an
excitation energy or temperature goes through a succession of decays,
the temperature of the produced fragments (assuming equilibration
before each decay) therefore also decreases as time proceeds.
In fig. 10, we display the average  temperature  $T_{av}$
of the produced  fragments  as a function of time when the
initial system $^{150}Sm$ has been prepared  at 
three  different excitation energies, namely,  $\epsilon^*
=13.5,$ 10.0  and 6.0 $MeV$. The temperature of the
fragments  is calculated from $T_{av}=(10 <\epsilon^*>)^{1/2}$ 
where $<\epsilon^*>$ is the ensemble averaged  excitation
energy per particle of the fragments
at any particular instant of time.
It is found that the higher the initial excitation energy of the system,
the faster is the cooling rate which is expected. An experimentalist
does not know a-priori  whether multifragmentation is a
one-step process (PM) or   is an outcome
of a sequence of binary decays. If one takes the fragmentation yields from
the SBD model as the 'experimental data' ,
it would be interesting to see the results
for the double ratio temperatures calculated with
the Albergo prescription as given by  eq. (15).
The double ratio temperatures so calculated for  the combinations
$(He-d)$,  $(He-t)$ 
$(Li-d)$ and $(Li-He)$ 
are
displayed in fig. 11.  One finds that except for  $(Li-d)$,
the temperatures are very  weakly dependent on  the initial excitation
energy and are very low ($\sim 3\> MeV$) even at the
highest excitation energy we study.
Such apparent temperatures  were obtained by Ma {\it et al}
\cite{ma1} in their Albergo-type analysis of the experimental
data in  $^{36}Ar+^{58}Ni$ collisions at 95$A\> MeV$.
For the $(Li-d)$ thermometer, the temperature
however rises  steadily  with initial excitation.
Thus the functional dependence of the temperature $T_r$ with
excitation energy obtained from  the SBD and PM
models are very different; the thermometers in the SBD model also
register   too low a temperature  compared to the PM model.

The kinetic  energy distribution of the fragments at the
end of the decay  process would   reflect the overall kinetic temperature
of the system. In the SBD model, 
since the system proceeds through a sequence of temperatures,
the kinetic energy distribution  reflects an
apparent temperature. In fig. 12,
this apparent temperature $T_{kin}$ is shown as a function of initial
excitation energy from the slope of the final energy
distributions of $p$, $d$, $^3He$ and $^4He$ produced
from $^{150}Sm$. The temperatures extracted from
the four distributions are not very different.
Closer inspection however shows that  except
for the  one for $^4He$, the 'Caloric curves' show broad
plateaux mimicing a liquid-gas phase transition. This arises
possibly  from the changing temperature scenario and a
complicated energy dependence of the fragment partial
widths for decay in the SBD model.

\section{CONCLUSIONS}

We have calculated the apparent temperatures from
several combinations of the double-ratio of isotope yields
in two different physical scenarios  in perspective; the
one-step prompt multifragmentation  and  the sequential
binary decay. In the PM model,  the
inclusion of  final state interaction  gives rise to nearly a constant
shift in the temperature $T_r$  calculated as a function of
excitation energy from the one obtained  from the Albergo 
prescription,  the shift being different for
different isotope combinations. The
effect of quantum statistics  on the apparent temperatures
is found  to be nominal; the effect
of  $\gamma$-feeding  is very substantial and is found to be
rather dramatic for the $(Be-He)$ thermometer. 
The presence of collective
flow  reduces the apparent temperature $T_r$  for a given total excitation
energy. Moreover, a soft plateau,  generally
seen in the caloric curves obtained for
the double-ratio temperatures becomes extended with inclusion of flow
energy. The import of our calculations
is that better contact  with the experimental data  can be achieved  
if one assumes  that the excitation
energy has a collective flow component in it.

One can not rule out  the sequential binary decay as a possible reaction
mechanism for the fragment yields,  particularly
at not too high excitation. This prompted us to study the
caloric curves  where the apparent temperatures 
$T_r$  are calculated from the fragment yields in the SBD  model, both  from
the double-ratios and slopes of the energy 
distributions of the fragments.  The double
ratio temperatures generally show  extended plateaux  but
no subsequent rise at  higher excitations; on the
other hand the caloric curves  calculated from the
slopes of the energy distributions  display  broad shoulders
with subsequent rise  at higher excitations  mimicing a first-order phase
transition. Since caloric curves obtained in  both
the PM model and the SBD model show apparent signatures
of a phase transition,   conclusion
regarding phase transition  in nuclear collision  requires  utmost
caution and  search for additional signatures is called for.

\newpage
\begin{center}
{\bf Figure Captions}
\end{center}

\begin{itemize}
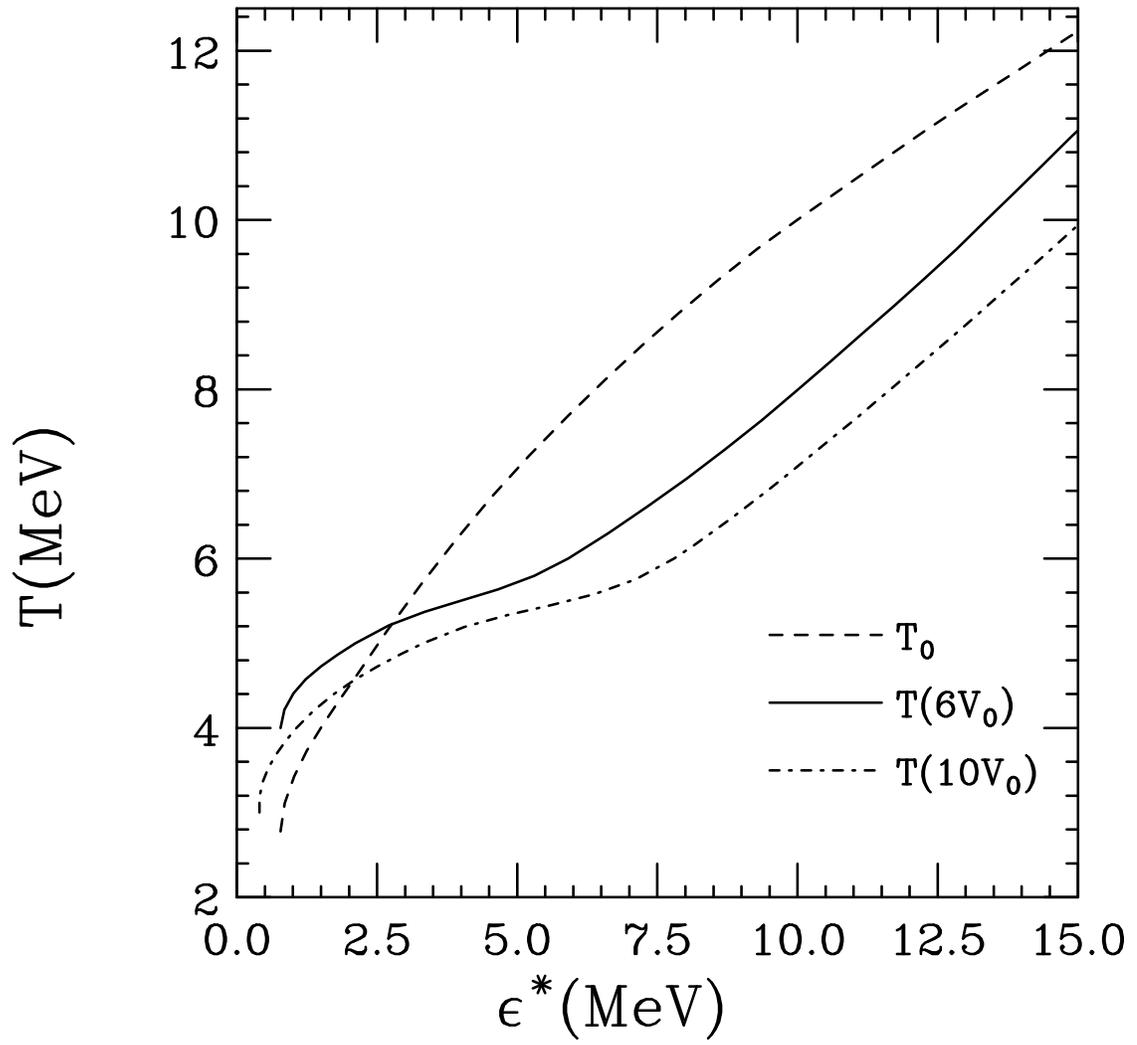
\item[Fig. 1]
The temperature of  the fragmenting system  $^{150}Sm$ as a 
function of $\epsilon^*$, the excitation energy per
nucleon.  The dashed line ($T_0$) corresponds to  the
temperature in the Fermi-gas approximation, the full and dot-dash
lines refer to temperatures at the freeze-out volumes taken to be 
$6V_0$ and $10V_0$ respectively.
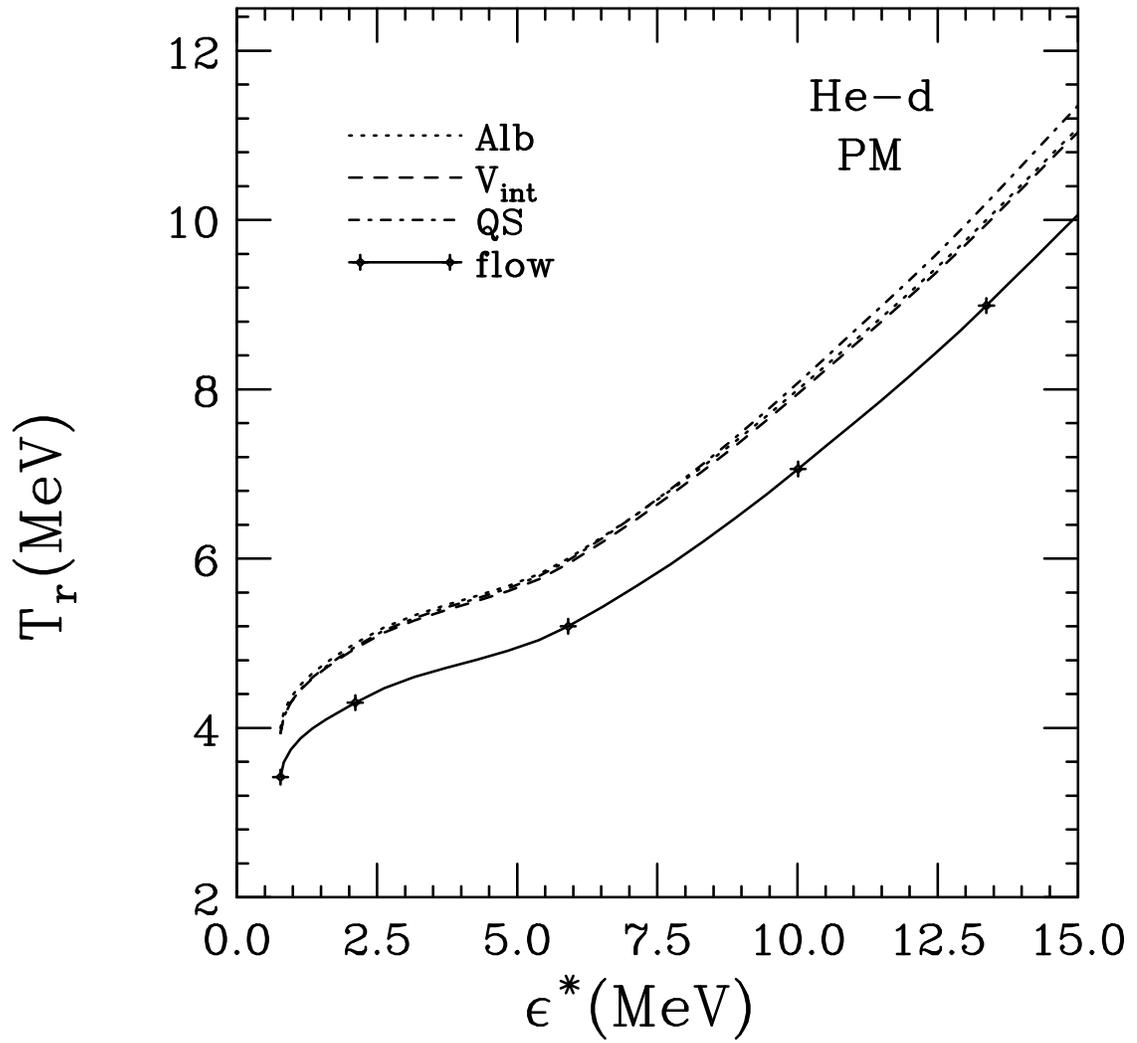
\item[Fig. 2]
The temperature $T_r$ from the double-ratio ($^{4}He/^{3}He)/(d/p)$
for the system $^{150}Sm$ as a function of $\epsilon^*$  in the 
prompt multifragmentation model. The dotted line refers to  $T_r$ 
obtained from Albergo prescription , the dashed, dot-dash and 
full line with crosses correspond to the temperatures with
subsequent progressive inclusion of  final state  interaction,
quantum statistics and flow energy, taken to be one-fourth of 
the total excitation.
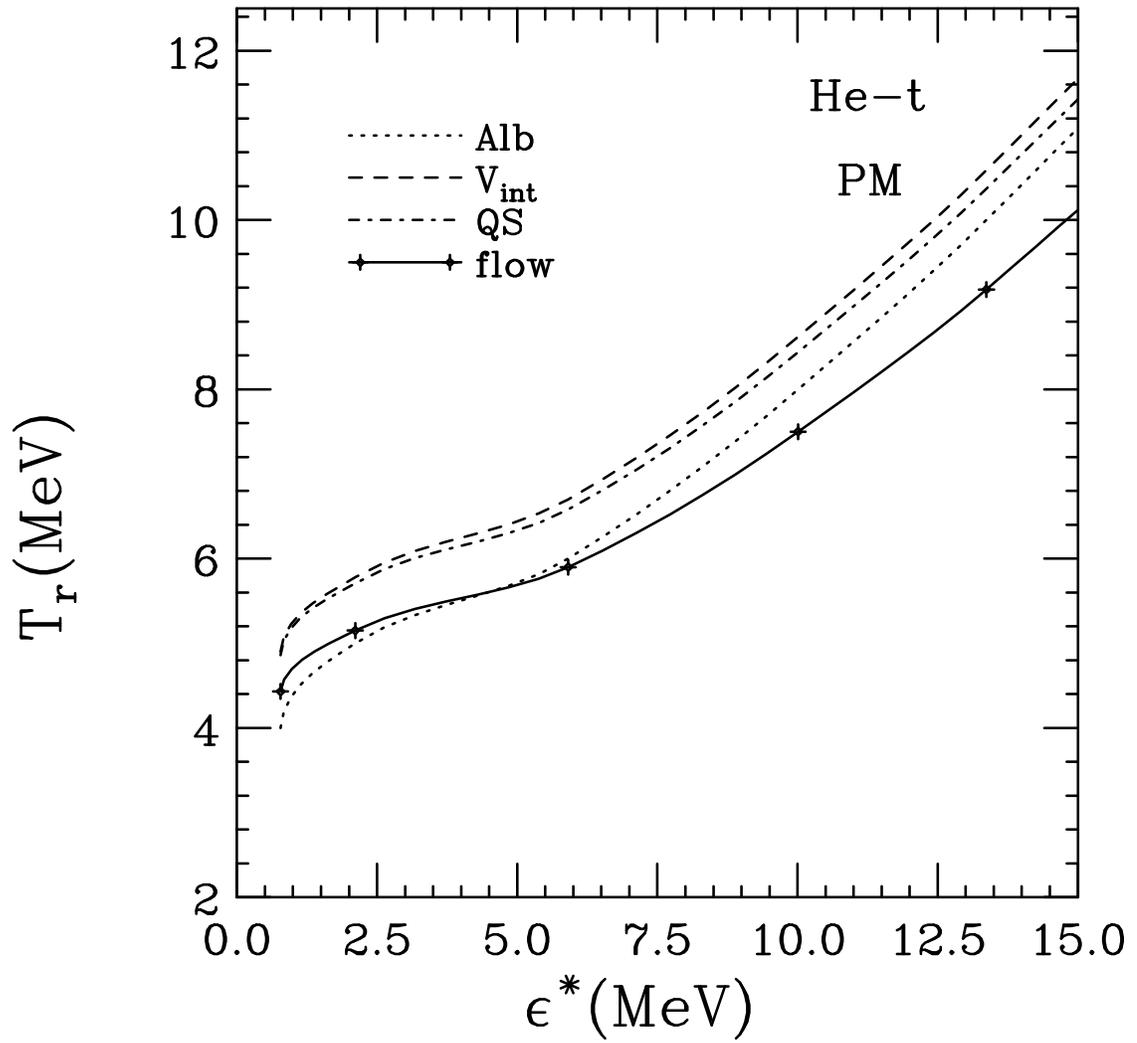
\item[Fig. 3]
Same as figure 2 for the ($^4He/^3He)/(t/d)$ thermometer.
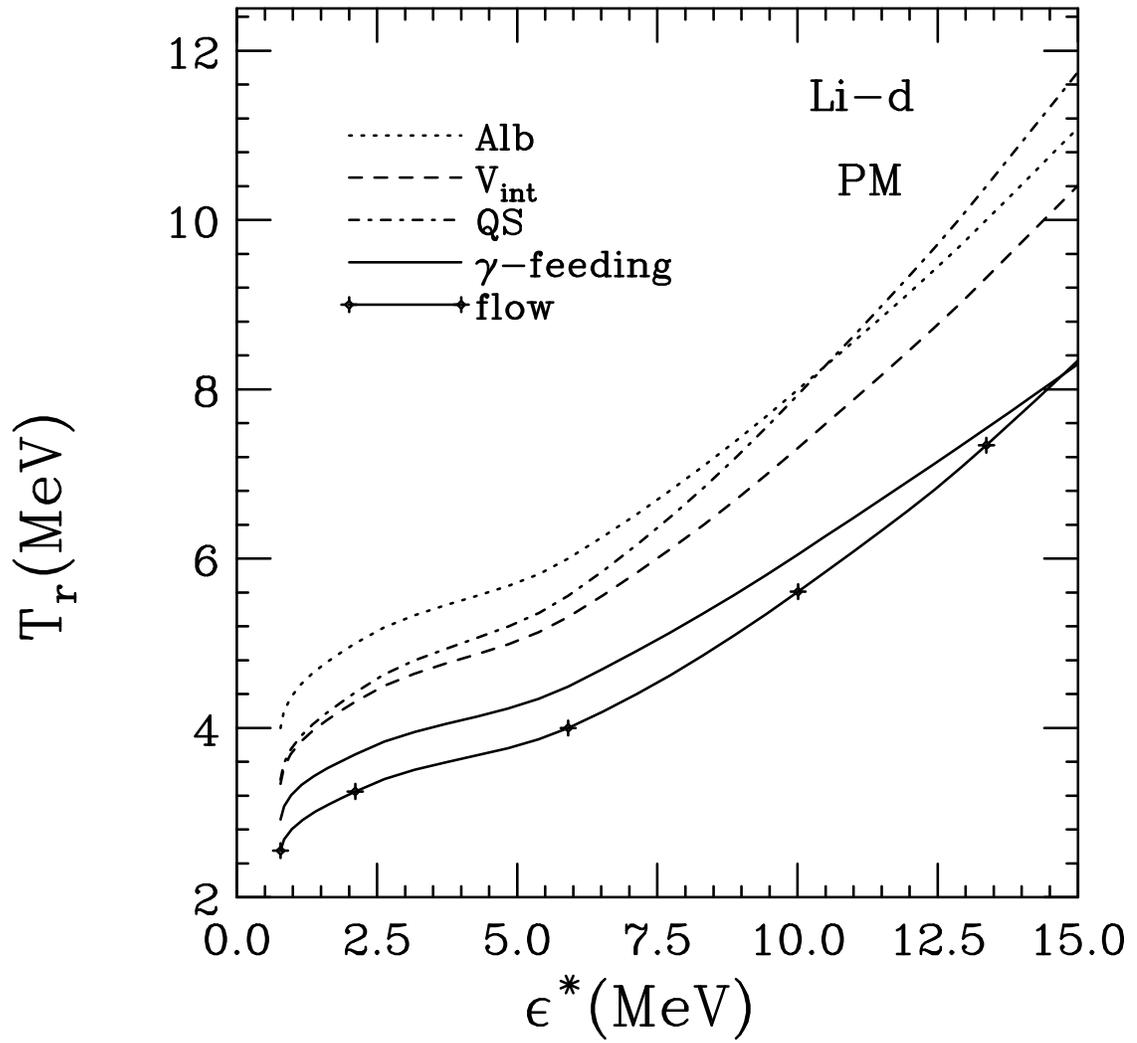
\item[Fig. 4]
The temperature $T_r$ from the yield ratio  $(^7Li/^6Li)/(d/p)$ in
the PM model.  The dotted line refers to 
the Albergo prescription, the dashed, dot-dash, full line
and line with crosses refer to  $T_r$ with subsequent  step by step 
inclusion of  final state interaction, quantum statistics,
$\gamma$-feeding and flow energy, taken to be one-fourth of the total
excitation.
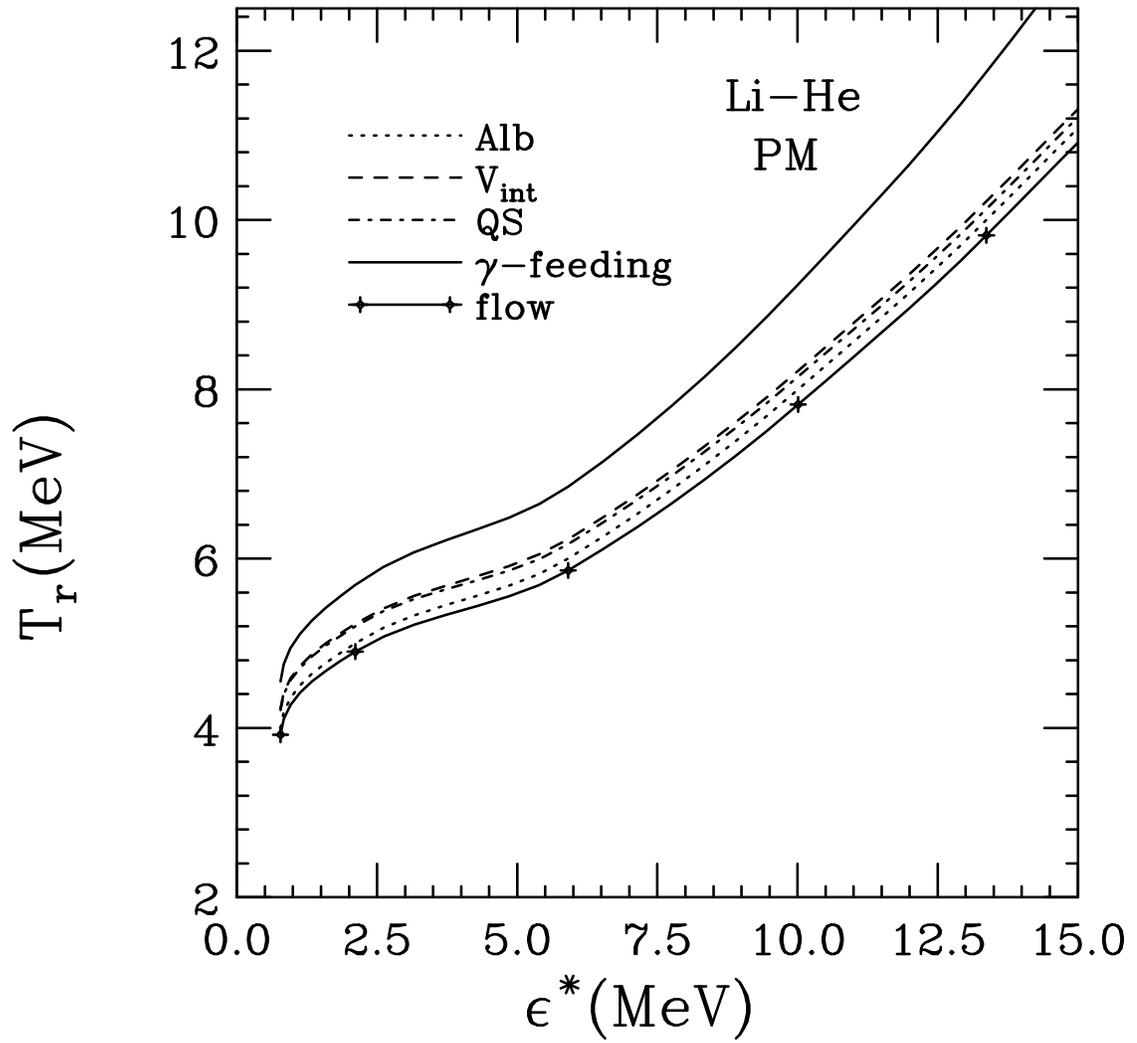
\item[Fig. 5]
Same as figure 4 for  the $(^7Li/^6Li)/(^4He/^3He)$
thermometer.
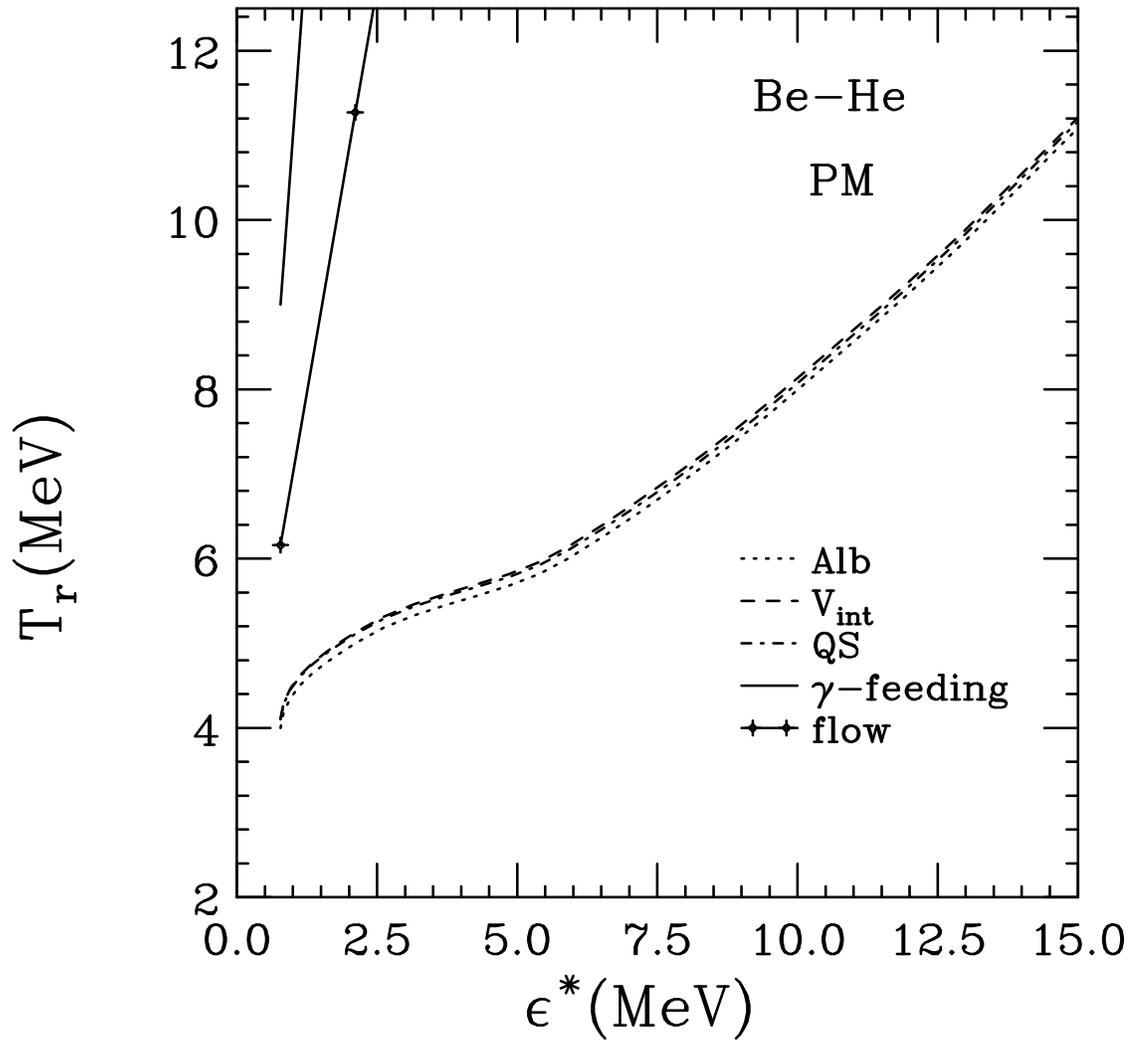
\item[Fig. 6]
Same as figure 4 for  the  $(^{10}Be/^9Be)/(^4He/^3He)$ 
thermometer.
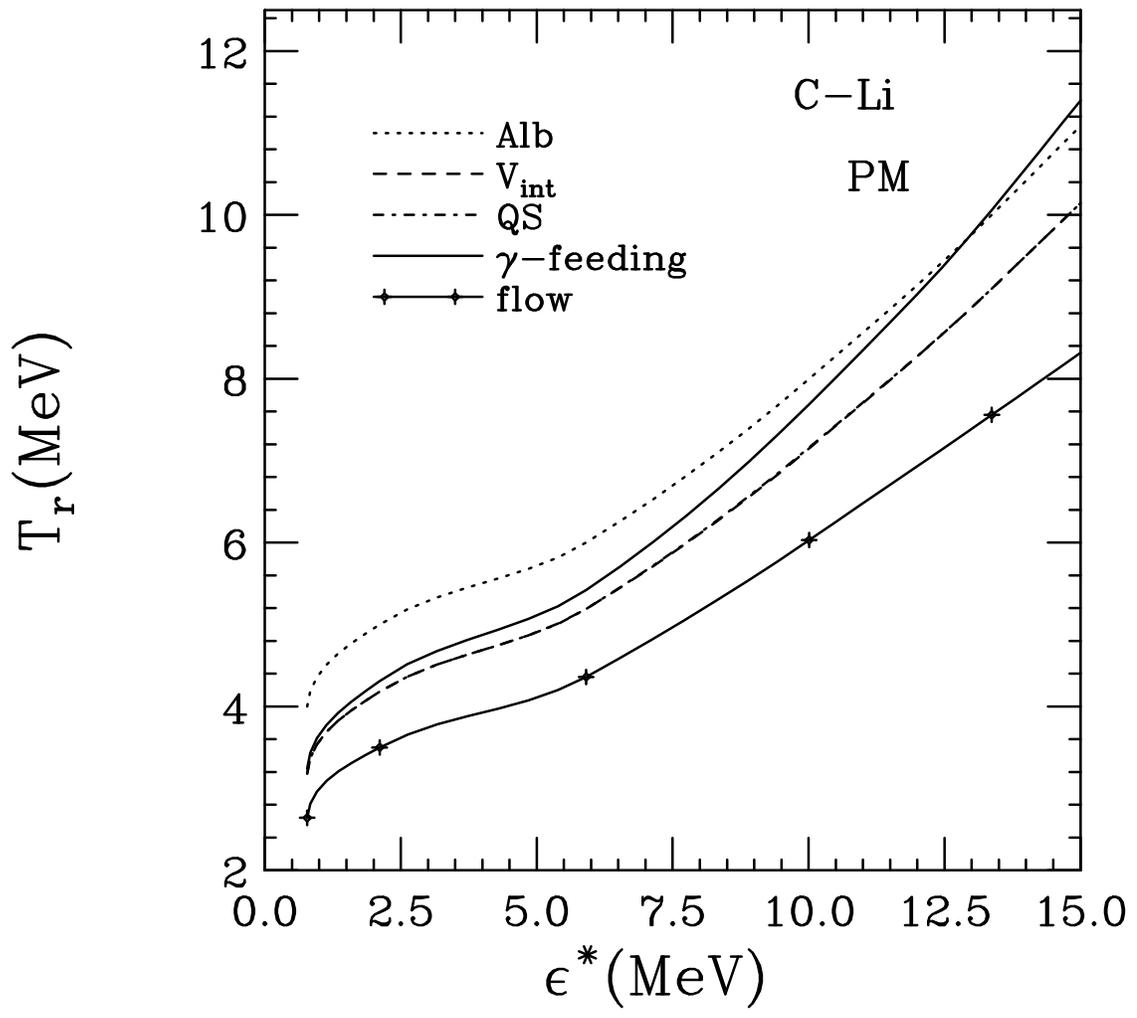
\item[Fig. 7]
Same as figure 4 for the $(^{13}C/^{12}C)/(^7Li/^6Li)$  thermometer. The
dashed and dot-dash curves can not be distinguished from each other.
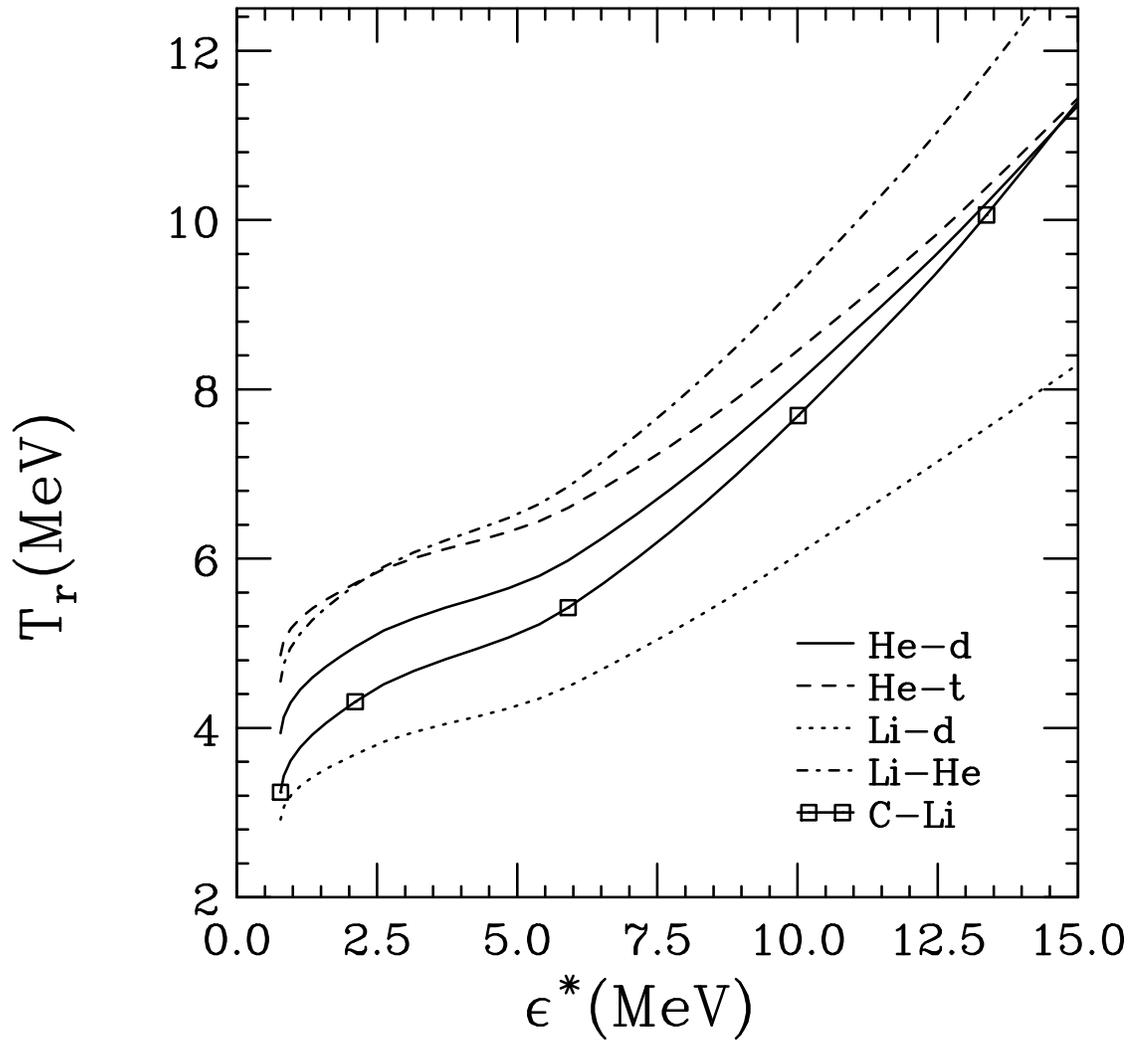
\item[Fig. 8]
The double-ratio temperatures $T_r$  obtained after
inclusion of  final state interaction, quantum statistics and
$\gamma$-feeding in the PM  model.
The fragmenting system is  $^{150}Sm$. The solid, dashed, dotted, dot-dash  
 and the full line with open squares refer to 
$(He-d)$, $(He-t)$, $(Li-d)$, $(Li-He)$ and
$(C-Li)$ thermometers respectively.
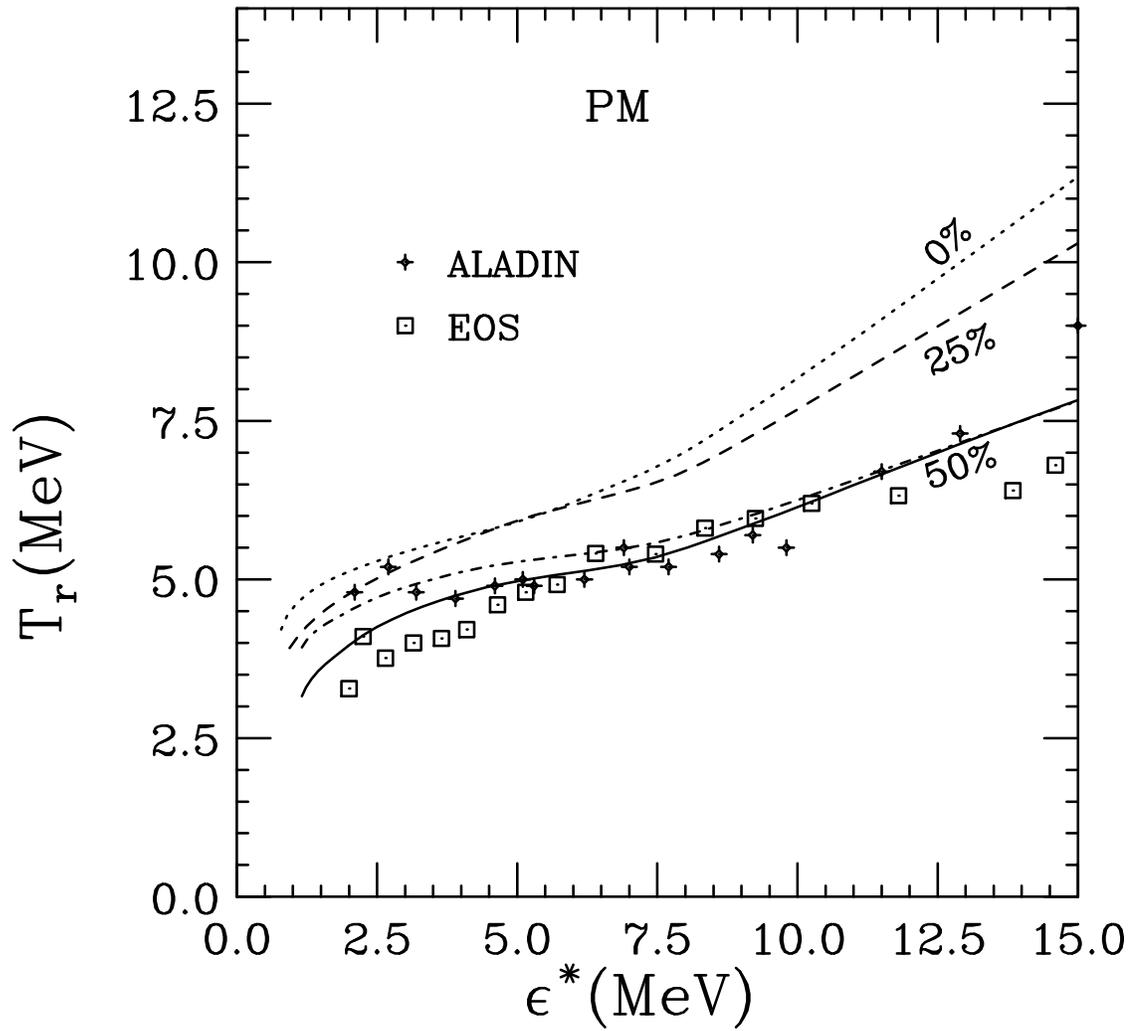
\item[Fig. 9]
The temperatures calculated with all effects  as discussed
in the text from the ($Li-He)$ and ($He-t$) 
thermometers in the PM model. The fragmenting system is $^{150}Sm$. 
The dotted,
dashed and full lines refer to  calculations for $(Li-He)$ temperatures 
with $0\%$, $25\%$ and $50\%$ of the total excitation as flow energy. 
The dot-dash line refers to the $(He-t)$ temperatures with flow energy as
$50\%$ of the total excitation.  
The crosses refer to ALADIN data and the open squares  
refer to the data from EOS collaboration.
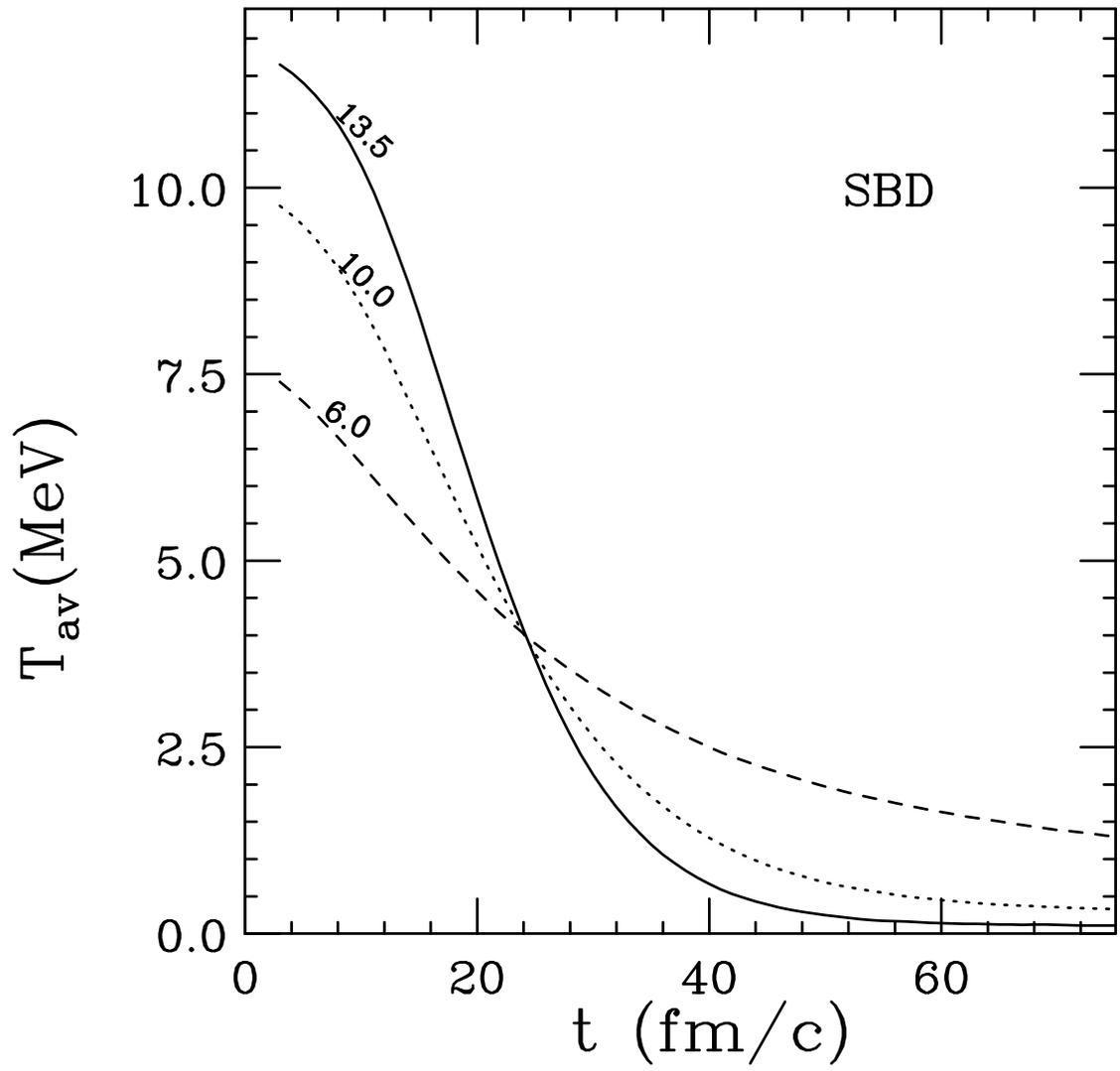
\item[Fig. 10]
Evolution of  the average temperature $T_{av}$  as a function
of time for the $^{150}Sm$ nucleus  prepared at excitations of 
13.5, 10.0 and 6.0 $MeV$ per nucleon respectively in the
SBD model.
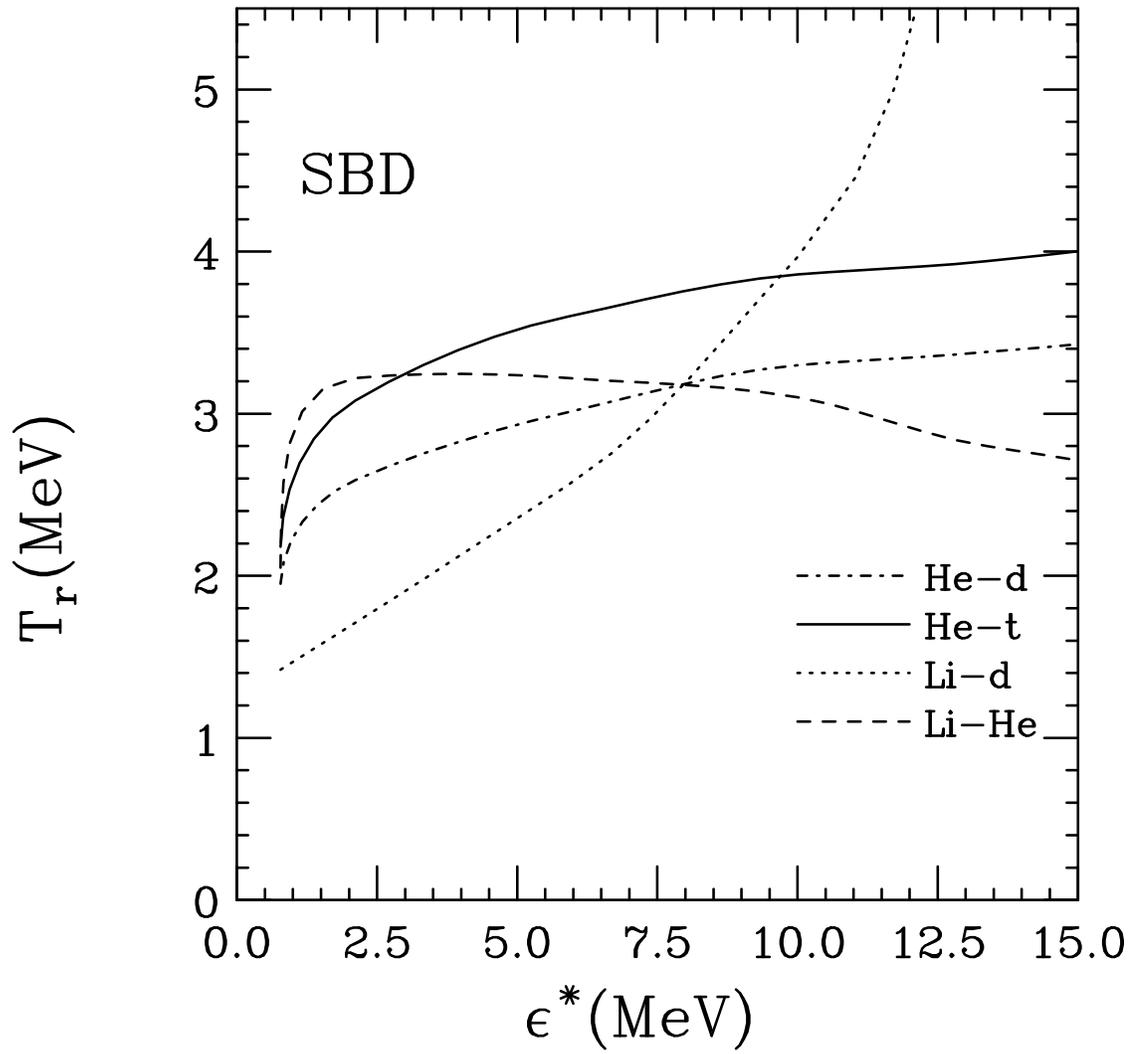
\item[Fig. 11]
The double-ratio  temperatures from the thermometers
$(He-d)$, $(He-t)$, $(Li-d)$ and $(Li-He)$ when the
fragments have been produced  in the SBD model from $^{150}Sm$.
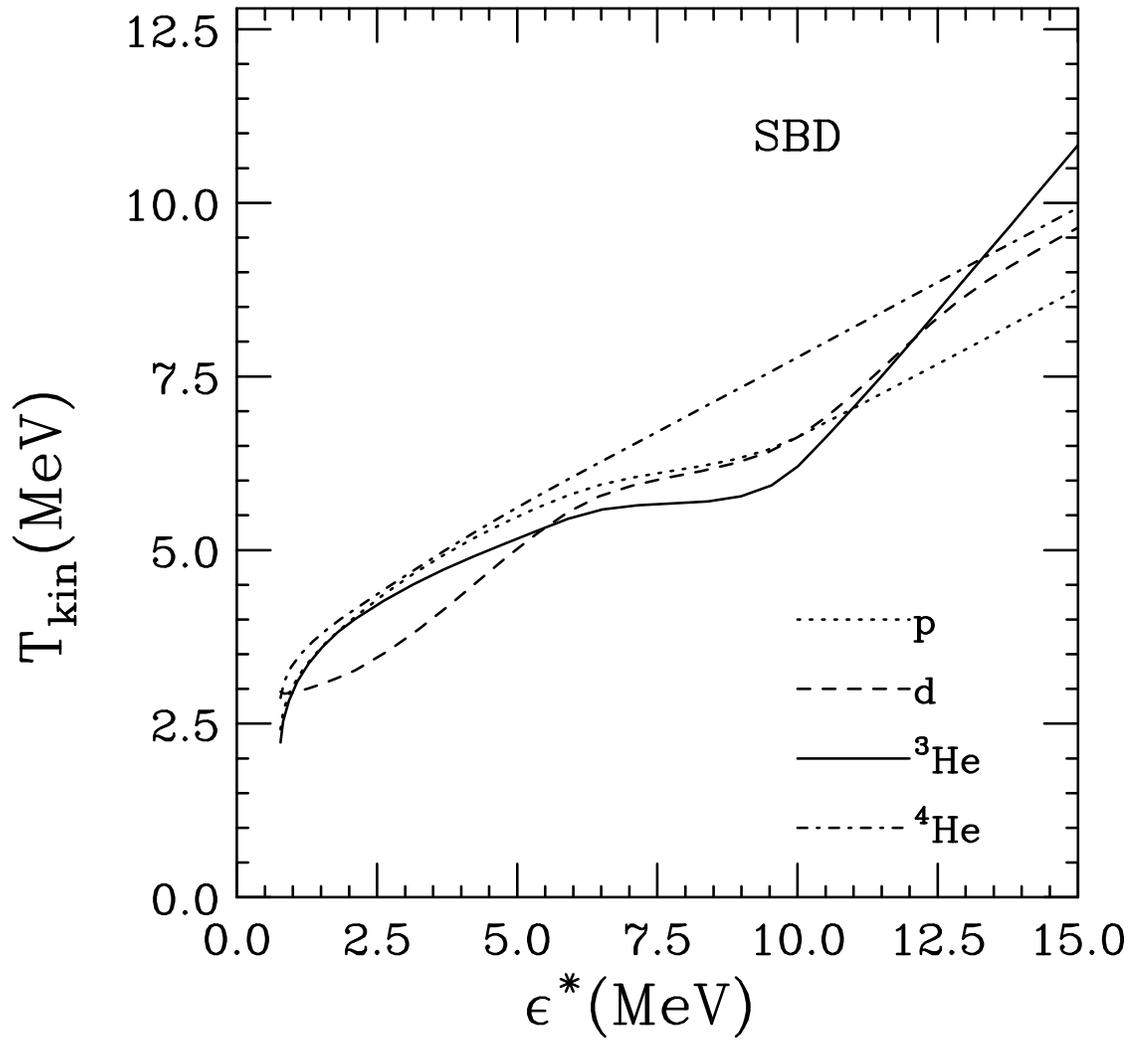
\item[Fig. 12]
The kinetic temperatures obtained from the energy distributions  of the fragments $p$,
$d$, $^3He$ and $^4He$ produced in the disassembly  of $^{150}Sm$
in the SBD model.
\end{itemize}

\end{document}